# Numerical and experimental studies of resonators with reduced resonant frequencies and small electrical sizes

T. Hao, J. Zhu, D.J. Edwards and C.J. Stevens

Department of Engineering Science, University of Oxford
Parks Road, Oxford, OX1 3DR, UK
Fax: + 44–1865273010; email: tong.hao@eng.ox.ac.uk

**Abstract**
Methods on reducing resonant frequencies and electrical sizes of resonators are reported in this paper. Theoretical and numerical analysis has been used and the results for the broadside-coupled resonators from both studies exhibit good agreement. Initial fabrication techniques are proposed and measurement results are compared with simulations. Further high resolution techniques have been envisaged to enhance the performance of the resonators. This class of small resonators with low resonant frequencies indicates a variety of applications in the design of microwave devices.

## 1. Introduction

Since late 1990s, intensive research has been carried out throughout the world after Pendry's work [1-2] on metamaterials which have both negative permittivity and permeability, where split ring resonators (SRR) were selected as resonant atoms to achieve negative permeabilities. More recent research is focused on increasing resonant frequencies into the optical region, and one of the highest frequencies achieved to date (~200THz) using the C-shaped ring structure is by Enkrich [3] where the current fabrication techniques reached their limit (50nm).

Heading an opposite direction, our research has been focussed on reducing resonant frequencies whilst also achieving small sizes by investigating different variations of C-shaped structures [4], followed Marqués work on the broadside-coupled SRR (BC-SRR) [5]. It has been verified that BC-SRR has a smaller electrical size than the traditional edge-coupled SRR (EC-SRR) (which is ~1/10), whilst VS-SRR can reduce it further. As is well known [2], a smaller electrical size is critical and must be satisfied before applying a continuous medium approach on metamaterial media, thus metamaterial devices with smaller physical sizes (in an order of millimeters) and lower frequencies (less than 1GHz) are of interest.

## 2. BC-SRR model

A model of BC-SRR is shown in Fig. 1. The upper ring and lower 180° oriented ring have identical dimensions, which are, the inner radius $r$, the width of rings $c$, the thickness of rings $h$, and the thickness of the insulator layer between rings $t$, the thickness of the substrate $d$, and the width of the gap $g$. Unlike edge-coupled resonators, the capacitance of BC-SRR is mainly contributed by the series capacitance of the upper and the lower rings, relatively less E field is concentrated in the split gap, thus the width of the gap is not the most important factor in determining the resonances. When fixing the thickness and dielectric constant of the insulator layer, the resonant frequency of the BC-SRR structure is mainly determined by the inner radius ($r$) and width ($c$) of the ring. As not pointed in [5] but could be calculated using [6], when fixing the oversize of the structure and varying $c$, one can achieve the lowest resonance if $c/r = 2/3$.

The normalised electrical size of BC-SRR can be written as $2(r+c)/\lambda_0$ [5], where $\lambda_0$ is the free space wavelength at resonance. Marqués et al showed that the normalised electrical size of the EC-SRR remains almost constant (~1/10) for small spacing between inner and outer rings. However, by tuning the thickness of insulator between rings $t$, and the dielectric constant of the insulator $\varepsilon$, smaller normalised electrical size can be achieved in the order of 1/300 at the resonance below 400MHz [7].

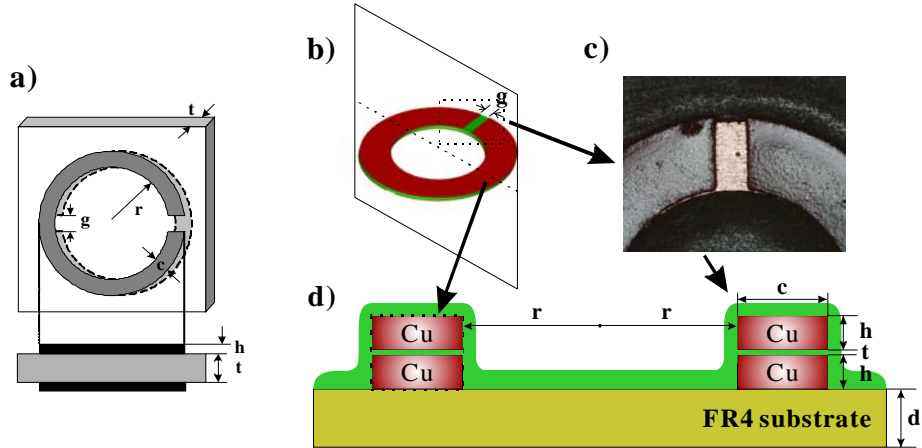

**Fig. 1** A model of BC-SRR. a): Marqués's BC-SRR model with thick substrate (~mm); b) BC-SRR model with very a thin insulator layer (~*um*) for simulations; c) plan view of BC-SRR (the gap part) made in the clean room; d) cross-section of BC-SRR

## 3. Results and discussion

Full-wave electromagnetic field analysis method is used in our simulation analysis. The commercial electromagnetic package used is MicroStripes® [8], which uses a 3D transmission line method (TLM) and takes as input a spatially discretised model of the object in order to determine the electromagnetic response of the model, in both time and frequency domains. A series of simulations have been conducted on varying the thickness and dielectric constant of the insulator layer for a relatively small resonator (*r*=0.9mm, *c*=0.6mm, and *g*=0.2mm). The results are summarised and compared with analytical calculations [5] in Table 1. The simulation results are generally larger than calculations since the theory doesn't take account of the small amount of capacitance in the split gaps.

**Table 1:** Comparison between simulated and theoretical predicted resonant frequencies in MHz (the BC-SRR dimensions: *r*=0.9mm, *c*=0.6mm, and *g*=0.2mm, h=30*um*, d=1.5mm)

| *t (um)* | Method | $\varepsilon_r$ | | | |
|---|---|---|---|---|---|
| | | 2.4 | 4 | 5 | 6 |
| 0.5 | Theory | 390.0 | 302.0 | N/A | N/A |
| | Simulation | 391.6 | 322.1 | | |
| 5 | Theory | 1227 | 951.0 | 851.0 | 777.0 |
| | Simulation | 1364 | 1028 | 920.0 | 873.1 |
| 10 | Theory | 1726 | 1340 | N/A | N/A |
| | Simulation | 1902 | 1465 | | |

As illustrated in Fig. 1(c), several resonator prototypes have been fabricated, measured and reported in [7]. The techniques adopted are the standard PCB etching technique, and cleanroom photolithography techniques employing S1805 photoresist and electroplating. Fig. 2 gives the resonances of several prototypes with various thicknesses of insulator layers.

It can be seen in Fig. 2 that the resonant frequency of the sample 1 (349.3MHz when *t*=0.5*um*, $\varepsilon_r$=2.4) has good agreement with the simulated (391.6MHz) and calculated (390.0MHz) ones in Table 1. The measured resonant frequencies of the fabricated samples did not show the constancy since it is very difficult to keep the thickness of the insulator layer constant in repeating the fabrication, where the resolution level was set to below 1*um*.

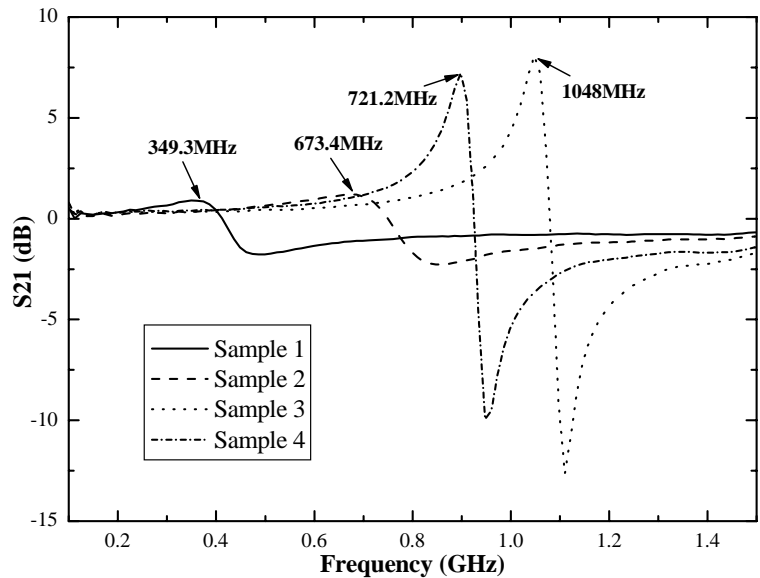

**Fig. 2** Resonances of prototype resonators with variable insulator layer thicknesses. Sample 1&2: 0.5um (Shipley 1805 (S1805) photoresist, relative dielectric constant=2.4, spinning speed=4000 rpm); Sample 3&4: (same material, 3000 rpm)

## 4. Conclusions

Numerical and theoretical calculations on BC-SRR structures with the insulator layer thickness below microns show a way to reduce the resonant frequencies, whilst increasing the dielectric constant of the layer will further lower the resonances. The reported techniques in fabricating BC-SRR with insulator layer thicknesses below 1*um* are not satisfactory, although a resonance below 400MHz has been achieved with the layer thickness of 0.5*um* and the normalised electrical size around 1/300. Different fabrication techniques with different insulator materials (e.g., nanolithography and pure PMMA [9]) can make the layer smoother and more uniform for the fabrication of 1D and 2D arrays of BC-SRR. And materials with higher dielectric constants (e.g., $TiO_2$ and $LiNiO_3$) are promising in making resonators at even lower resonances.

## References
[1]. J. B. Pendry, A. J. Holden, W. J. Stewart, and I. Youngs, "Extremely low frequency plasmons in metallic mesostructures," *Physical Review Letters*, vol. 76, pp. 4773-6, 1996.
[2]. J. B. Pendry, A. J. Holden, D. J. Robbins, and W. J. Stewart, "Magnetism from conductors and enhanced nonlinear phenomena," *IEEE Transactions on Microwave Theory and Techniques*, vol. 47, pp. 2075-84, 1999.
[3]. Enkrich-C, Wegener-M, Linden-S, Burger-S, Zschiedrich-L, Schmidt-F, Zhou-Jf, Koschny-T, and Soukoulis-Cm, "Magnetic metamaterials at Telecommunication and visible frequencies," *Physical Review Letters*, vol. 95, pp. 203901/1-4, 2005.
[4]. T. Hao, C. J. Stevens, and D. J. Edwards, "Optimisation of metamaterials by Q factor," *Electronics Letters*, vol. 41, pp. 653-4, 2005.
[5]. R. Marques, F. Mesa, J. Martel, and F. Medina, "Comparative analysis of edge- and broadside- coupled split ring resonators for metamaterial design - theory and experiments," *IEEE Transactions on Antennas and Propagation*, vol. 51(10) pt. 2, pp. 2572-81, 2003.
[6]. http://james.eii.us.es/srrCalculator/
[7]. T. Hao, C. J. Stevens, and D. J. Edwards, "A realisation of resonators with small physical and electrical sizes," *Electronics Letters* (submitted).
[8]. www.microstripes.com
[9]. Martin-C, Rius-G, Borrise-X, and Perez-Murano-F, "Nanolithography on thin layers of PMMA using atomic force microscopy," *Nanotechnology*, vol. 16, pp. 1016-22, 2005.